\documentstyle [eqsecnum,preprint,aps,epsf]{revtex}
\begin {document}

\def\alphaw{\alpha_{\rm w}}
\def\Ncs{N_{\rm cs}}
\def\tr{{\rm tr}}
\def\Tr{{\rm Tr}}
\def\v#1{{\bf #1}}
\let\mat=\hat
\def\PiL{\Pi_{\rm L}}
\def\PiT{\Pi_{\rm T}}
\def\Ab{\v A_{\rm b}}
\def\Re{{\rm Re}}
\def\Im{{\rm Im}}
\def\A{\v A}
\def\x{\v x}
\def\p{\v p}
\def\q{\v q}

\def\ehat{\hat{\v e}}
\def\DR{D_{\rm R}}
\def\half{{\textstyle{1\over2}}}
\def\G{{\cal G}}

\makeatletter
%
%
%
\def\footnotesize{\@setsize\footnotesize{10.0pt}\xpt\@xpt
\abovedisplayskip 10\p@ plus2\p@ minus5\p@
\belowdisplayskip \abovedisplayskip
\abovedisplayshortskip  \z@ plus3\p@
\belowdisplayshortskip  6\p@ plus3\p@ minus3\p@
\def\@listi{\leftmargin\leftmargini
\topsep 6\p@ plus2\p@ minus2\p@\parsep 3\p@ plus2\p@ minus\p@
\itemsep \parsep}}
%
%
\long\def\@makefntext#1{\parindent 5pt\hsize\columnwidth\parskip0pt\relax
\def\strut{\vrule width0pt height0pt depth1.75pt\relax}%
$\m@th^{\@thefnmark}$#1}
%
%
\long\def\@makecaption#1#2{%
\setbox\@testboxa\hbox{\outertabfalse %
\reset@font\footnotesize\rm#1\penalty10000\hskip.5em plus.2em\ignorespaces#2}%
\setbox\@testboxb\vbox{\hsize\@capwidth
\ifdim\wd\@testboxa<\hsize %
\hbox to\hsize{\hfil\box\@testboxa\hfil}%
\else %
\footnotesize
\parindent \ifpreprintsty 1.5em \else 1em \fi
\unhbox\@testboxa\par
\fi
}%
\box\@testboxb
} %
%
%
\global\firstfigfalse
\global\firsttabfalse
%
%
\def\tabular{\let\@halignto\@empty\@tabular}
\def\endtabular{\crcr\egroup\egroup $\egroup}
\expandafter \def\csname tabular*\endcsname #1{\def\@halignto{to#1}\@tabular}
\expandafter \let \csname endtabular*\endcsname = \endtabular
\def\@tabular{\leavevmode \hbox \bgroup $\let\@acol\@tabacol
   \let\@classz\@tabclassz
   \let\@classiv\@tabclassiv \let\\\@tabularcr\@tabarray}
\def\endtable{%
\global\tableonfalse\global\outertabfalse
{\let\protect\relax\small\vskip2pt\@tablenotes\par}\xdef\@tablenotes{}%
\egroup
}%
\makeatother

\preprint {UW/PT-96-19}

\title  {The hot baryon violation rate is $O(\alphaw^5 T^4)$}

\author {Peter Arnold, Dam Son, and Laurence G.~Yaffe}

\address
    {%
    Department of Physics,
    University of Washington,
    Seattle, Washington 98195
    }%
\date {September 1996}

\maketitle
\vskip -20pt

\begin {abstract}%
{%
{%
The rate per unit volume for anomalous electroweak baryon number
violation at high temperatures, in the symmetric phase, has been
estimated in the literature to be $O(\alphaw^4 T^4)$ based on simple
scaling arguments.  We argue that damping effects in the plasma suppress
the rate by an extra power of $\alphaw$ to give $O(\alphaw^5 T^4)$.
We show how to understand this effect in a variety of ways
ranging from an effective description of the long-distance modes responsible
for baryon number violation, to a microscopic picture of the short-distance
modes responsible for damping.  In particular, we resolve an old controversy
as to whether damping effects are relevant.
Finally, we argue that similar damping effects should
occur in numerical simulations of the rate in classical thermal field
theory on a spatial lattice, and we point out a potential problem with
simulations in the literature that have not found such an effect.
}%
\ifpreprintsty
\thispagestyle {empty}
\newpage
\thispagestyle {empty}
\vbox to \vsize
    {%
    \vfill \baselineskip .28cm \par \font\tinyrm=cmr7 \tinyrm \noindent
    \narrower
    This report was prepared as an account of work sponsored by the
    United States Government.
    Neither the United States nor the United States Department of Energy,
    nor any of their employees, nor any of their contractors,
    subcontractors, or their employees, makes any warranty,
    express or implied, or assumes any legal liability or
    responsibility for the product or process disclosed,
    or represents that its use would not infringe privately-owned rights.
    By acceptance of this article, the publisher and/or recipient
    acknowledges the U.S.~Government's right to retain a non-exclusive,
    royalty-free license in and to any copyright covering this paper.%
    }%
\fi
}%
\end {abstract}


\section {Introduction}

Anomalous baryon number (B) violation in the
hot, symmetric phase%
\footnote{
   Here and throughout, we use the term ``symmetric phase'' loosely since,
   depending on the details of the Higgs sector, there many not be any
   sharp boundary between the symmetric and ``symmetry-broken'' phases
   of the theory \cite{kajantie,elitzur}.
   By symmetric phase we shall mean
   temperatures high enough that the magnetic correlation length
   is $O(1/g^2 T)$ and determined by non-perturbative dynamics.
}
of electroweak theory occurs through the creation of
non-perturbative, nearly static, magnetic configurations with spatial extent
of order $(g^2 T)^{-1}$, where $g$ is the electroweak SU(2) coupling.
(This is reviewed below.)
It has been widely assumed that $g^2 T$ must then be the only
scale relevant to the problem,
so that the baryon number violation rate per unit volume,
by dimensional analysis, must be $O\bigl((g^2 T)^4\bigr)$.
We shall argue that damping effects in the plasma cause the time
scale for transitions through these configurations to be $O(g^4 T)$
instead of $O(g^2 T)$, and therefore the rate is
$O\bigl((g^2 T)^3 (g^4 T)\bigr) = O(\alphaw^5 T^4)$.

The possible importance of damping effects was pointed out many years ago
in ref.~\cite{arnold&mclerran}
in the context of B violation in the symmetry-broken phase of the theory.
The effect was controversial, and another analysis of the
problem \cite{khlebnikov} claimed that damping plays no role.
In the intervening years,
many people have privately expressed the concern that damping might affect
the symmetric phase rate, and we make no claim to be the only, or even the
first, people to think of it.
The purpose of this paper is simply to
resolve the controversy, elucidate the physics involved, and put to paper
the result that the rate is $O(\alphaw^5 T^4)$.
Many of the individual parts of this paper will be reviews of
various aspects of thermal physics already familiar to some readers,
but we believe their synthesis in this discussion of the
baryon number violation rate is original.

In the remainder of this introduction we briefly review the conventional
picture of anomalous transitions in the symmetric phase and the
standard estimate of the rate.
Then we give a quick but formal estimate of the effects of damping
that closely follows the broken-phase discussion of damping
in ref.~\cite{arnold&mclerran}.

The body of the paper is devoted to providing a number of different ways
in which to understand the physics underlying this result.
Section II begins with an analysis of the power spectrum of gauge
field fluctuations based on the fluctuation-dissipation theorem 
and argues that it is fluctuations with frequency of order $g^4 T$
and spatial extent $(g^2 T)^{-1}$ which have sufficiently large
amplitude to generate non-perturbative transitions.
This is followed by a real-time, finite-temperature diagrammatic analysis 
which demonstrates that perturbation theory breaks down in exactly
this domain due to the non-linear interactions of low-frequency,
low-momentum components of the non-Abelian magnetic field.
Then, in section III we sketch the microphysical origin of damping
and turn to one of the traditional methods of understanding the
transition rate: by computing the rate at which the system crosses
the ``ridge'' separating classically inequivalent vacua.
We show that a single {\it net} transition from the neighborhood of
one valley to the next actually involves $O(1/g^2)$ back-and-forth
crossings of this ridge, and so standard methods which count individual
ridge-crossings overcount the true rate by a factor of $1/g^2$---precisely
the suppression due to damping.
In the final section, we discuss how damping effects should also
play a roll in the topological transition rate for classical thermal field
theory on a spatial lattice.
This appears inconsistent with the results of numerical simulations
\cite{ambjorn}, which show no sign of damping effects.
We explain why the simulations done so far may fail to measure the
true topological transition rate.

This multiplicity of approaches in not intended to be a case of
several poor arguments ``substituting'' for one good one.
We believe that {\em any one} of the viewpoints discussed below
provides compelling evidence that thermal damping is responsible
for suppressing the high-temperature electroweak baryon violation rate
by one power of $\alphaw$ relative to the naive order $\alphaw^4 T^4$
estimate.
Our goal is simply to examine the phenomena from multiple perspectives
and understand the connections between them.


\begin {figure}
\vbox
   {%
   \begin {center}
      \leavevmode
      
      \epsfbox [150 260 500 530] {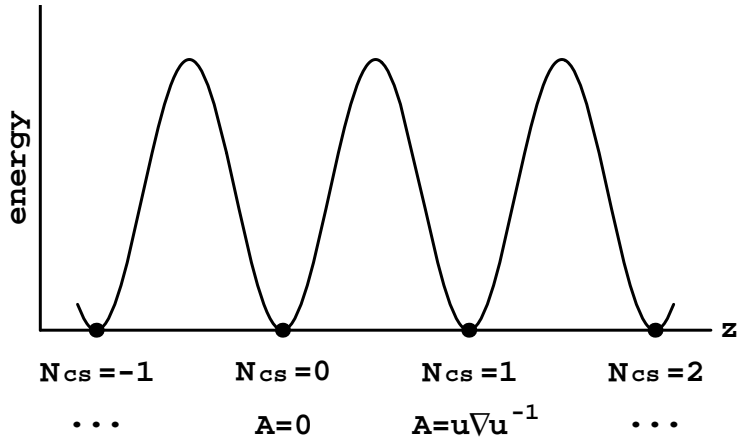}
   \end {center}
   \caption
       {%
       A schematic representation of
       the (bosonic) potential energy along a particular direction (labeled
       $z$) in field space, corresponding to topologically non-trivial
       transitions
       between vacua.
       \label{fig:ridge1}
       }%
   }%
\end {figure}

\subsection{Review of the standard picture}

Fig.~\ref{fig:ridge1} is the standard visual aid for thinking about anomalous
transitions.  Consider the theory in Hamiltonian formalism or in
$A_0 = 0$ gauge, where the degrees of freedom are $\A(\x)$ and
the conjugate momenta are $\v E(\x) = -\partial\A(x) / \partial t$.
The horizontal axis represents one particular direction in
the infinite-dimensional space of gauge configurations
$\A(\x)$.  The minima represent the vacuum $\A = 0$ and
large gauge transformations of it, labeled by their Chern-Simons
number $\Ncs$.
The vertical axis represents the potential energy of the configurations.%
\footnote{
   More precisely, the non-fermionic contribution to the potential energy.
   When a transition is made, there will also be the perturbative energy cost
   of the fermions created by that transition.
}
Whenever a transition is made from the
neighborhood of one minimum to another
(which we call a {\it topological} transition),
the electroweak anomaly
causes baryon number to be violated by an amount proportional to
$\Delta \Ncs$:%
\footnote{
   There are also other directions in configuration space along which
   $\Ncs$ changes that have nothing to do with the vacuum structure of
   the theory and exist even in U(1) theories.  These directions are
   not ultimately relevant to baryon number violation.  To make the issue
   more precise, imagine starting with a cold system with some baryon
   number, heating it up for a time to make anomalous transitions possible,
   and then quickly cooling it.
   The system will cool into the nearest vacuum state shown in
   fig.~\ref{fig:ridge1}.
   So the net change in baryon number in this example depends on whether
   the system has made a transition from the neighborhood of one vacuum state
   to the next, not simply on whether Chern-Simon number has temporarily
   changed due to an excursion in some irrelevant direction.
   See \cite{christ} for a more detailed discussion.
   }
\begin {equation}
   \Delta B \propto \Delta\Ncs =
   - {g^2 \over 16\pi^2} \int d^4x \> \tr \, F \tilde F \,.
\label{eq:anomaly}
\end {equation}
That said, we shall now ignore the fermions and, for simplicity,
focus on the rate for topological transitions in the pure bosonic theory.
And, since we are interested in the symmetric phase of the theory, we shall
generally ignore the Higgs sector as well and focus on pure SU(2) gauge
theory.%
\footnote
    {%
    Except in the immediate vicinity of the electroweak phase transition
    (or crossover),
    fermions or Higgs fields merely provide additional ``hard'' thermal
    excitations in the symmetric phase and do not affect any of the
    following discussion in a substantive fashion.
    }

Topological transitions can occur at a significant rate if thermal
fluctuations have enough energy to get over the top of the energy barrier
separating neighboring vacua.
The most important parameter determining the energy of the barrier is the
spatial extent $R$ of the configurations depicted in fig.~\ref{fig:ridge1}.
In order to generate an $O(1)$ change in Chern-Simons number
(or baryon number), Eq.~(\ref{eq:anomaly}) plus dimensional analysis
implies that a gauge field of spatial extent $R$ must have a field strength%
\footnote
    {%
    We are using the conventional gauge field normalization in
    which no factor of $1/g^2$ multiplies the kinetic terms in the action
    and perturbative fluctuations are $O(1)$ in amplitude.
    }
of order $(g R^2)^{-1}$.
Consequently, the energy of these configurations
on the potential energy barrier
will be $O((g^2 R)^{-1})$,
while the gauge field amplitude itself (smoothed over the scale $R$)
is $O((g R)^{-1})$.

\begin {figure}
\setlength\unitlength{1 in}
\vbox
   {%
   \begin {center}
   \begin {picture}(5,3)(0.1,0)
      \put(0.0,0.4){
        \leavevmode
        
        \epsfbox [150 200 500 500] {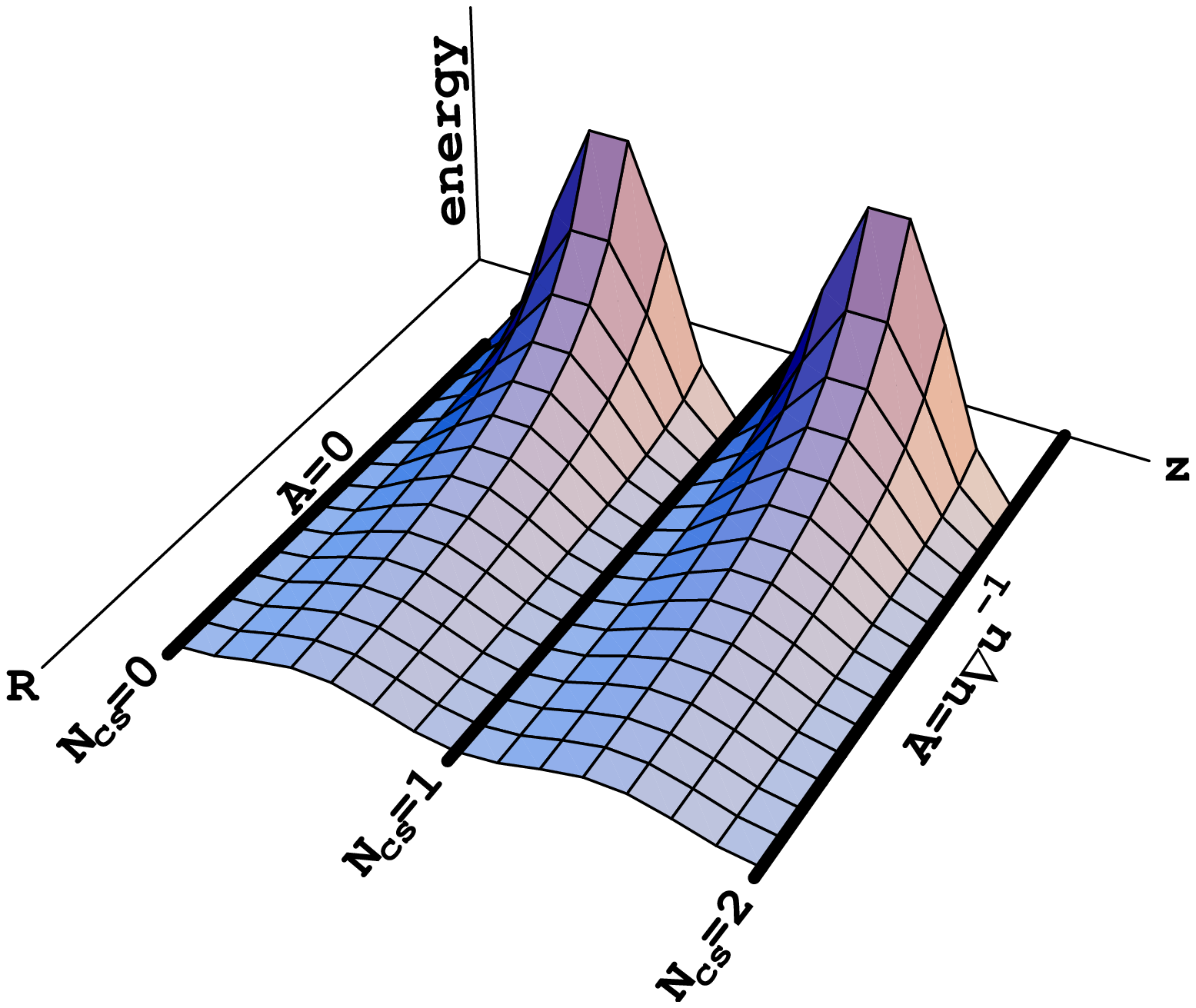}
      } 
      \put(3.0,0.4){
        \leavevmode
        
        \epsfbox [150 200 500 500] {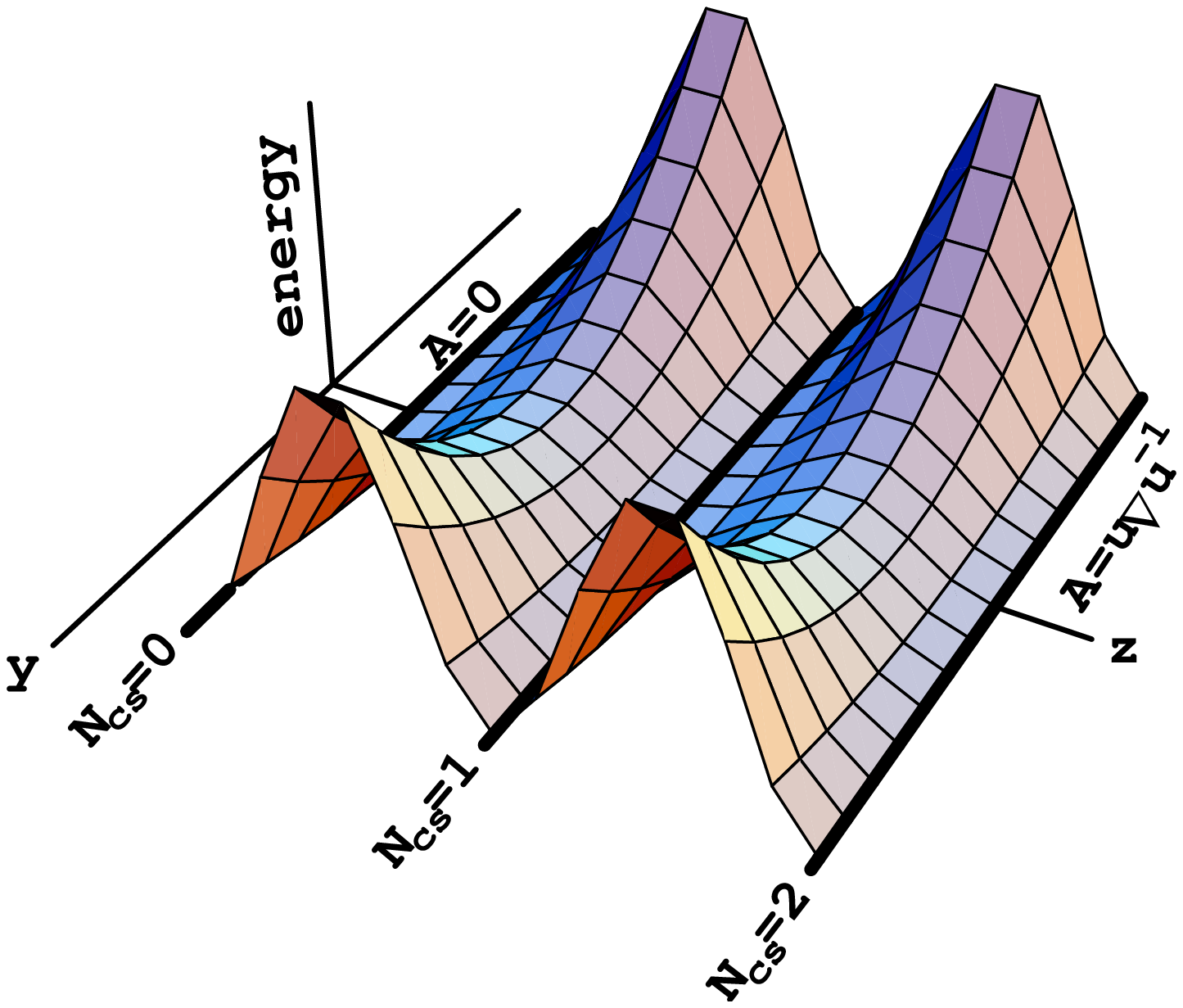}
      }
      \put(1.0,0.1){(a)}
      \put(4.0,0.1){(b)}
   \end {picture}
   \end {center}
   \caption
       {%
       The same as fig.~\protect\ref{fig:ridge1} but supplemented by an
       extra dimension of configuration space corresponding to
       (a) spatial size $R$ of the configurations, and (b) some other
       generic direction $y$, such as a particular mode of momentum $\sim T$.
       \label{fig:ridge2}
       }%
   }%
\end {figure}

So a more representative picture of the configuration space is
provided by fig.~\ref{fig:ridge2}a:
the energy barrier is not a single point but a ridge which
becomes arbitrarily low if arbitrarily large configurations are used
to cross it.
The smallest $R$ for which non-perturbative thermal transitions through
such configurations are not significantly Boltzmann-suppressed is
$R \sim 1/g^2 T$.
For the same reason, this is also the scale where perturbation theory
breaks down in the hot plasma.

$R \sim 1/g^2 T$ is in fact the dominant spatial scale
for topological transitions.
One argument is entropy: there are fewer
ways to cross the barrier with larger configurations than smaller ones.
Another argument is due to magnetic confinement in the hot plasma.
Ignore $\v E = -\dot \A$ at the energy barrier for the moment, so that
the configuration is purely magnetic.  Though static non-Abelian
electric forces are no longer confining at high temperature,
static magnetic forces are.%
\footnote{
  The standard way to see this is to consider the expectation of very large
  {\it spatial} Wilson loops at high temperature.  These can be evaluated in
  the Euclidean formulation of finite-temperature gauge theory, where
  Euclidean time has a very small period $\beta$ at high temperature.
  By dimensional reduction \protect\cite{3d reduction},
  this is equivalent to understanding the
  behavior of large Wilson loops in three-dimensional Euclidean theory.
  But three-dimensional SU(2) gauge theory is confining.
}
The confinement scale is just the spatial scale
of non-perturbative physics, $1/g^2 T$.


Now consider the transition rate.  Suppose that a
configuration of size $R$ on the energy barrier ridge
of fig.~\ref{fig:ridge2}a is produced in the plasma.
What would be the time scale for its decay?  The unstable modes of
the configuration will be associated with momenta of order $1/R$.
If the configuration were decaying at zero temperature, the time scale
for decay would then be $O(1/R)$.  So the standard estimate is that
there is one unsuppressed transition per volume $R^3$ per time $R$,
giving the rate $\Gamma \sim R^{-4} \sim (g^2 T)^4$.  As we shall see,
the flaw in this estimate is that the configuration is not decaying at
zero temperature but instead in interacting with the other excited modes of
the plasma.

We should emphasize that the picture of fig.~\ref{fig:ridge2}a
is still
incomplete because we have shown only two degrees of freedom in
configuration space.  There are an infinite number of other degrees
of freedom in which the potential turns up so that, for fixed $R$,
the energy barrier looks like a saddle as depicted in fig.~\ref{fig:ridge2}b.
The energy ridges discussed before are really hypersurfaces through
configuration space, separating valleys that contain the different
vacua.  One can formally define these hypersurfaces as follows
\cite{khlebnikov}:
For each configuration
$\A(\x)$, follow the steepest descent path away from it in the
infinite-dimensional analog of fig.~\ref{fig:ridge2}---that is, follow
(minus) the gradient of the potential energy.
For generic configurations,
one will eventually approach one of the classical vacua.
The barrier ridge hypersurface which separates the classical vacua
is the exceptional hypersurface which doesn't flow to a vacuum configuration.
We shall return to this picture of the ridges in detail later,
when we take up the old controversy about whether damping can affect
the transition rate.


\subsection {A quick estimate of damping}

Consider again
the decay of a configuration $\Ab (\x)$
of size $1/g^2 T$ on the energy barrier.
As the system passes through $\Ab$, one may analyze the motion by
linearizing the equation of motion.  Expanding
$\A(\x,t) = \Ab(\x) + \delta\A(\x, t)$,
the equations have the form
\begin {equation}
   \partial_t^2 \, \delta\A(\x) = \v J(\x)
       - \mat K^2 \, \delta\A(\x) + O(\delta\A^2) \,,
\end {equation}
where $\v J$ is the gradient of the potential along the ridge and
$\mat K^2$ is the potential energy curvature operator
with a negative eigenvalue $-\kappa^2$
of order $(g^2 T)^2$
corresponding to decay away from the ridge.
For the moment, let's focus on that one eigenmode, reducing our equation to
\begin {equation}
   \partial_t^2 \, \delta A = \kappa^2 \, \delta A + O(\delta A^2) \,.
\label{eq:decay0}
\end {equation}
The solution to this equation grows exponentially on a time scale of $1/g^2T$.

\begin {figure}
\vbox
   {%
   \bigskip
   \begin {center}
      \leavevmode
      
      \epsfbox [150 320 500 450] {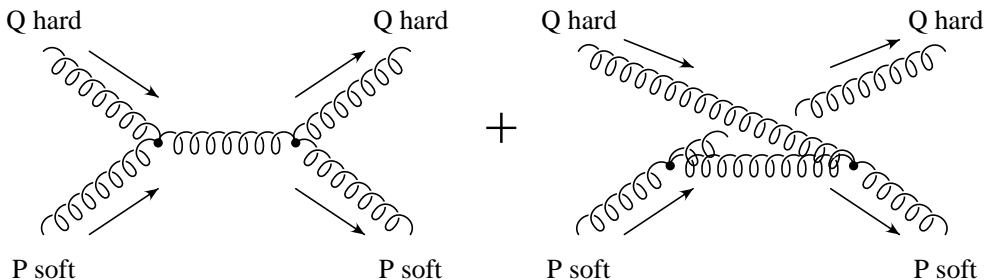}
   \end {center}
   \caption
       {%
       The soft-mode self-energy generated by forward scattering off of
       hard modes.  The external hard lines are on-shell.
       \label{fig:pi}
       }%
   }%
\end {figure}

Now consider the interaction of this decay with typical thermal excitations
of the plasma.
The transition rate involves physics at soft energies $g^2 T$ that 
are small
compared to the hard energies $T$ of typical particles in the system.
The simplest way to analyze the problem is to consider an effective theory
for the soft modes of the theory, where the physics
of the hard modes has been integrated out.  In the context
of (\ref{eq:decay0}), we need to know the effective $(\delta A)^2$ interaction
generated by integrating out the hard modes.  The soft-mode self-energy
generated by the hard modes is dominated by the processes shown in
fig.~\ref{fig:pi} and is known as the hard thermal loop approximation to the
self-energy.%
\footnote{
   The hard thermal loop approximation also generates corrections
   to cubic and higher couplings of multiple soft particles
   \cite{hard thermal loops}.
   These will be needed in a complete quantitative analysis,
   but they do not affect any of our simple estimates.
}

There is a reasonably simple formula for the resulting self-energy
$\Pi(\omega,p)$ in the limit $\omega,p \ll T$ \cite{pi},
but for our purposes it will only be important to know its
qualitative behavior in terms of the ratio $\eta \equiv \omega/p$:
\begin {mathletters}%
\label{eq:PiLT}%
\begin {eqnarray}
   \PiL &=& \cases{
        g^2 T^2 \, (O(1) - i \, O(\eta)),   & $\eta \ll 1$ \,, \cr
        g^2 T^2 \, O(1),                    & $\eta \ge 1$ \,,\cr
   }_{\strut}
\\
   \PiT &=& \cases{
        g^2 T^2 \, (O(\eta^2) - i \, O(\eta)), & $\eta \ll 1$ \,, \cr
        g^2 T^2 \, O(1),                       & $\eta \ge 1$ \,,\cr
   }^{\strut}
\end {eqnarray}%
\end {mathletters}%
where $\PiL$ and $\PiT$ are the longitudinal and transverse parts of
the (retarded) one-loop self-energy.
To distinguish them, it is convenient to momentarily switch to covariant
gauge%
\footnote{
  For $A_0=0$ gauge, ${\cal P}_{\rm L}$ of (\ref{eq:PL}) is restricted to
  spatial indices and is then $-P_0^2/P^2$ times the projection
  operator $\bar{\cal P}_{\rm L} = p^i p^j/|\v p|^2$.  There is a
  compensating factor of $-P^2/P_0^2$ in the longitudinal piece of the
  free propagator in $A_0=0$ gauge:
  $iD^{(0)} = P^{-2} \left({\cal P}_{\rm T}
     - (P^2/P_0^2) \bar{\cal P}_{\rm L}\right)$.
}
(rather than $A_0 = 0$ gauge), where the
longitudinal and transverse projection operators are
\begin {mathletters}%
\begin {eqnarray}
   {\cal P}_{\rm L}^{\mu\nu} &=& - {P^2\over |{\v p}|^2}
       \left(g^{\mu 0} - {P^\mu P^0 \over P^2} \right)
       \left(g^{0 \nu} - {P^0 P^\nu \over P^2} \right) \,,
\label{eq:PL}
\\
   {\cal P}_{\rm T}^{ij} &=& \delta^{ij} - {p^ip^j\over|\p|^2} \,,
\end {eqnarray}%
\end {mathletters}%
where $P^\mu \equiv (\omega,\p)$ and ${\cal P}_T$ has only spatial components.
The $O(g^2 T^2)$ behavior of $\PiL(0,\p)$ reflects the Debye screening
of static electric fields at distances of $1/gT$.
The vanishing of $\PiT(0,\p)$ reflects the absence of similar
screening for static magnetic fields.
The $O(g^2 T^2)$ behavior of both $\PiT(\omega,0)$ and $\PiL(\omega,0)$
reflects the $O(g T)$ mass gap for propagating plasma waves.

\begin {figure}
\vbox
   {%
   \begin {center}
      \leavevmode
      
      \epsfbox [150 350 500 460] {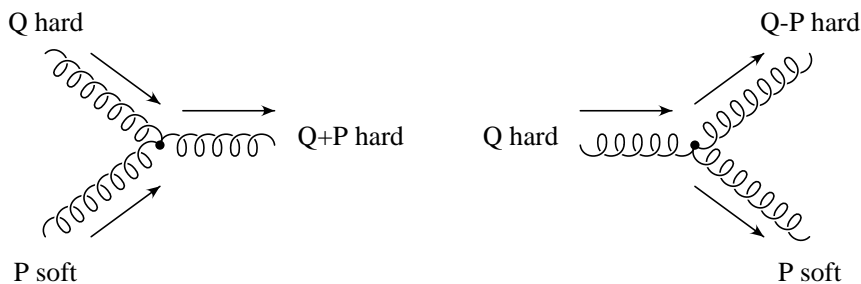}
   \end {center}
   \caption
       {%
       Absorption or emission of soft modes by hard modes when $\omega < p$,
       contributing to $\Im\Pi$.
       \label{fig:impi}
       }%
   }%
\end {figure}

The imaginary part of $\PiT$ arises from
absorption or emission due to scattering off hard particles
in the thermal bath, as depicted in fig.~\ref{fig:impi}.
By kinematics, these processes
occur only when $\omega < p$.
Now modify eq.~(\ref{eq:decay0}) to
include the self-energy (and Fourier transform to frequency space),
\begin {equation}
   - \omega^2 \delta A = (\kappa^2 - \Pi(\omega))\delta A \,.
\label{eq:dA0}
\end {equation}
Given that $\kappa \sim p \sim g^2 T$,
the self-energy $\Pi(\omega)$ will stabilize
the unstable mode unless we focus on transverse modes and take $\eta \ll 1$.
We may then approximate
\begin {equation}
   \Pi \sim i g^2 T^2 \, \eta \sim i \, \omega T \,,
\end {equation}
so that the solution to the linearized Eq.~(\ref{eq:dA0}) is
\begin {equation}
   \omega \sim i g^4 T \,.
\end {equation}
The decay time is therefore of order $1/g^4 T$,
or $1/g^2$ slower than assumed in the standard non-dissipative estimate.

\section {Some alternative views}

\subsection {Analysis in terms of spectral density}

We now consider another formal way to see that $\omega \sim g^4 T$ is
the appropriate
frequency
scale for topological transitions.
Start by considering
the power spectrum $\rho_A(\omega,\p)$ of gauge-field
fluctuations $\A(\omega,\p)$ in the plasma:
\begin {equation}
   \left\langle \A^i(\omega,\p)^* \A^j(\omega',\p') \right\rangle
   \equiv
   \rho_A^{ij}(\omega,\p) \, (2\pi)^4 \delta(\omega{-}\omega')
               \, \delta^3(\p{-}\p')
   \,.
\label{eq:rho def}
\end {equation}
This may be related to the retarded propagator
\begin {eqnarray}
   i \DR^{ij}(t{-}t',\x{-}\x')
   &=& \left\langle [ \A^i(t,\x)^*, \A^j(t',\x') ]\right\rangle \,
       \theta(t{-}t')
\nonumber\\
   &=& Z^{-1} \, \Tr \, e^{-\beta H} [\A^i(t,\x)^*, \A^j(t',\x')] \,
       \theta(t{-}t')
\label{eq:Ddef}
\end {eqnarray}
by the fluctuation-dissipation theorem:%
\footnote{
  The theorem follows from inserting a complete set of energy eigenstates
  in (\protect\ref{eq:Ddef}), Fourier transforming, and then taking
  the imaginary part.
  See, for example, sec.~31 of ref.~\cite{fetter&walecka}.
}
\begin {equation}
   \rho_A(\omega,\p)
   = - 2 (n_\omega+1) \, \Im \DR(\omega,\p)
   = - 2 (n_\omega+1) \, \Im \left(
        1 \over (\omega+i\epsilon)^2 - \p^2 - \Pi(\omega,\p)
     \right)
   \,,
\label{eq:imD}
\end {equation}
where $n_\omega = 1/(e^{\beta\omega}-1)$
is the Bose distribution function.

\begin {figure}
\vbox
   {%
   \begin {center}
      \leavevmode
      
      \epsfbox [150 300 500 530] {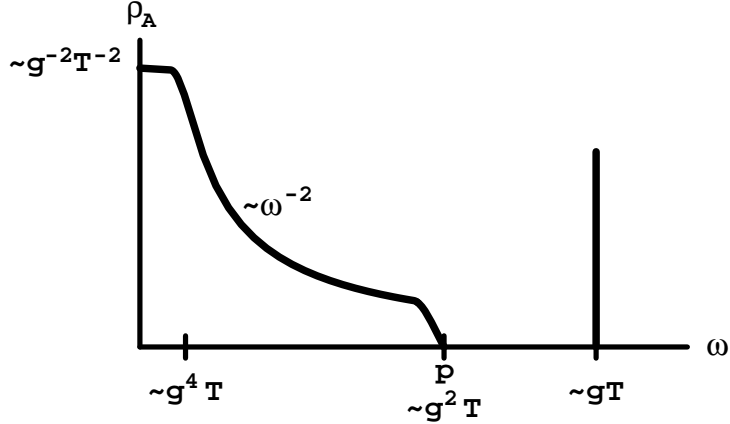}
   \end {center}
   \bigskip
   \caption
       {%
       A schematic plot of the power $\rho_A(\omega)$ of gauge-field
       fluctuations with $p \sim g^2 T$.
       The $\delta$-function spike at $\omega \sim gT$ corresponds to
       propagating plasmons.
       \label{fig:power}
       }%
   }%
\end {figure}

Now, using this relation, consider which frequencies of $A$
will have enough power
to generate topological transitions.
As reviewed in section 1,
this requires fields with spatial extent $R \sim 1/g^2 T$
and amplitude $A \sim 1/gR \sim gT$.
Consider the right-hand side of (\ref{eq:imD}) and use the
hard thermal loop approximation (\ref{eq:PiLT}) for $\Pi$ in
\begin {equation}
   \Im\DR(\omega,\v p) = {\Im\Pi \over (\omega^2-\p^2-\Re\Pi)^2 + (\Im\Pi)^2}
            ~-~ \delta\bigl(\omega^2 - O(g^2 T^2)\bigr)
\end {equation}
The last term corresponds to propagating plasma waves.%
\footnote{
   It is only a $\delta$-function in the leading-order approximation to
   $\Pi$.  The smearing of the $\delta$-function by the plasmon width,
   and other features of the spectrum such as the two-plasmon cut, will
   not be important for the order-of-magnitude power estimates we make
   below.
}
For $p \sim g^2 T$,
the integrated power given by the right-hand side of (\ref{eq:imD}) from
all frequencies of order $\omega$ is of order
\begin {equation}
   \omega \rho_A \sim \cases{
       1/g^2T        \,,    &    $\omega \sim g T$ \,;               \cr
       1/\omega      \,,    &    $g^4 T \lesssim \omega \le p$  \,;  \cr
       \omega/g^8T^2 \,,    &    $\omega \lesssim g^4 T$ \,.         \cr
   }
\label{eq:power}
\end {equation}
For $\omega \ll gT$, it is dominated by transverse fluctuations.
A schematic plot of the power $\rho_A(\omega)$ is shown in
fig.~\ref{fig:power}.
From (\ref{eq:rho def}), the power in a particular soft mode with
$R \sim 1/g^2 T$, and frequency of order $\omega$, will produce
fluctuations with amplitude
\begin {equation}
   A
   \sim (p^3 \omega \rho_A)^{1/2}
   \sim \cases{
       g^2 T           \,,    &    $\omega \sim g T$ \,;               \cr
       g^3 \sqrt{T^3/\omega}
                       \,,    &    $g^4 T \lesssim \omega \le p$  \,;  \cr
       g^{-1} \sqrt{\omega T}
                       \,,    &    $\omega \lesssim g^4 T$ \,.         \cr
   }
\end {equation}
Hence,
non-perturbative amplitudes $A \sim 1/gR \sim gT$ are generated when
$\omega \sim g^4 T$.

In section 3, we shall see how to get this same power spectrum by
considering the microphysical details of the behavior of the hard
degrees of freedom, and we will translate the power spectrum into a
qualitative discussion of what the actual real-time trajectories of the soft
degrees of freedom look like.



\subsection{Estimate from Feynman diagrams}

\begin {figure}
\vbox
   {%
   \begin {center}
      \leavevmode
      
      \epsfbox [150 320 500 490] {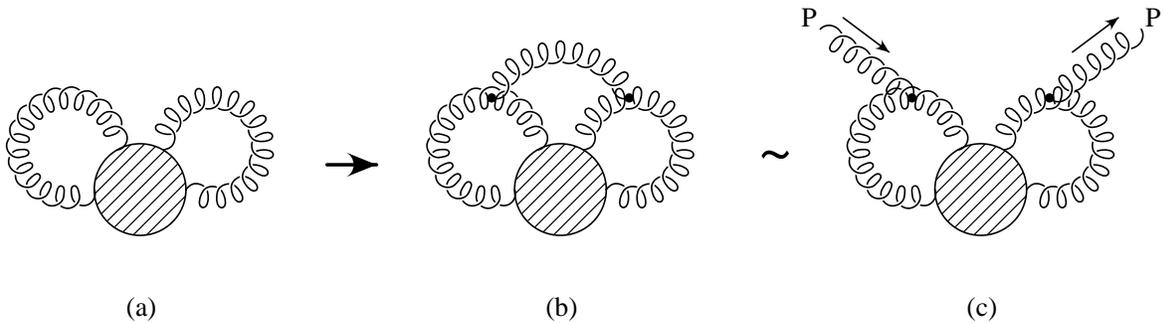}
   \end {center}
   \caption
       {%
       Adding a loop (a$\to$b) to a Feynman diagram.  (c) represents the
       thermal contribution to the new loop, corresponding to forward
       scattering off a particle in the thermal bath.
       \label{fig:loop1}
       }%
   }%
\end {figure}

A pictorial way to represent the previous argument is to ask
when diagrammatic perturbation theory breaks down in the effective
theory of the soft modes, since we need non-perturbative effects for
a topological transition.  For simplicity, let us focus on transverse modes
with $\omega \ll p$, for which one can
ignore $\Re\Pi$.  Consider adding a loop to a Feynman diagram,
as shown in fig.~\ref{fig:loop1},
and consider the thermal contribution to that loop,
shown in fig.~\ref{fig:loop1}c.
The cost of adding the loop is order $(gp)^2$ from the
new vertices, $p^{-4}$ from the two new uncut propagators, and
$d^3 p\, d\omega \> \rho_A(\omega, \p)$ for the phase space probability of
finding the new soft particle in fig.~\ref{fig:loop1}c, where
$\rho_A \sim n_\omega\, \Im\DR$.  Integrating over frequencies of
order $\omega$ and momenta of order $p \sim g^2 T$, the cost of adding
a loop is of order
\begin {equation}
   (gp)^2 \times p^{-4} \times p^3\omega\, {T\over\omega}\, \Im\DR
   \sim g^4 T^2\, \Im\DR
   \sim \cases{
       g^4 T / \omega \,,   & $g^4 T \lesssim \omega \le p$   \,, \cr
       \omega / g^4 T \,,   & $\omega \lesssim g^4 T$         \,. \cr
   }
\end {equation}
So the loop expansion parameter is $O(1)$ for $\omega \sim g^4 T$.

\subsection {Interacting magnetic fields}

\begin {figure}
\vbox
   {%
   \begin {center}
      \leavevmode
      
      \epsfbox [150 290 500 600] {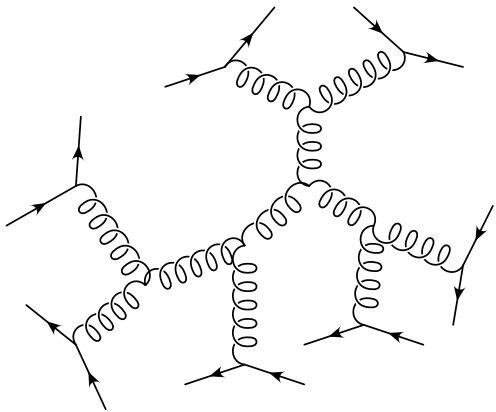}
   \end {center}
   \caption
       {%
       A microphysical picture of the important interactions for
       topological transitions.
       \label{fig:loop2}
       }%
   }%
\end {figure}

Instead of considering Feynman diagrams in the effective theory
(in which the self-energy has been resummed), we can
get some insight into the origin of the transition time scale by recasting
the above argument in terms of diagrams in the original, microscopic theory.
The key to the estimates above was the behavior of $\Im\PiT$, which
arises from the interactions shown in fig.~\ref{fig:impi}.
The origin of the dominant
contribution arising from multiple interactions of the form of
fig.~\ref{fig:loop1}c is interactions such as those shown in
fig.~\ref{fig:loop2}.  The straight
lines represent hard thermal particles; the wavy lines represent
virtual magnetic quanta with $(\omega,p) \sim (g^4T,g^2T)$ that are
absorbed or emitted. 

The cost to the {\it rate} of adding a new
interaction with a new hard particle
is $(p^{-4})^2$ for new propagators, $(gp)^2$ for a new soft vertex,
$(gT)^2$ for a new hard vertex, $O(1)$ for hard particle Bose or Fermi
factors, and
\begin {equation}
   {d^3 q \over 2\omega_q} {d^3 q'\over2\omega_{q'}}
     \; \theta(|\q {-} \q'| \mathrel{\lesssim} p)
     \; \theta(|\omega_q {-}\omega_{q'}| \mathrel{\lesssim} \omega)
   ~\sim~ T \omega p^2
\end {equation}
for phase space.
The total cost, by this estimate, is
\begin {equation}
   (p^{-4})^2 \times (gp)^2 \times (gT)^2 \times O(1) \times T \omega p^2
   \sim {\omega \over g^4 T}
\end {equation}
and is indeed $O(1)$ for $\omega \sim g^4 T$.  For larger $\omega$, where
this microscopic
perturbation theory appears to go completely wild, one must consider the
effect of resumming the effects of $\Im\Pi$ into the soft propagators and
return to our previous estimates.  The advantage to having considered
fig.~\ref{fig:loop2}
is simply that it gives us a physical picture for the origin
of topological transitions in the plasma.  The soft, virtual lines emitted from the
hard particles simply correspond to the low-frequency and low-momentum
components of the magnetic fields produced by the movement of those
particles.  These are not propagating electromagnetic waves but simply
the magnetic fields carried by all moving charged particles.  Topological
transitions are then created by the non-linear interactions of these
fields.


\section {Microphysical picture of barrier crossing}

The normal approach used
to calculate the topological transition rate in the
symmetry-broken phase, and to estimate it in the symmetric phase,
is based on calculating
the rate, in equilibrium, at which the system crosses the
energy barrier hypersurface separating the classical
vacua \cite{arnold&mclerran,khlebnikov}.
If one assumes that
each such crossing is associated with a net transition of the system from
the neighborhood of one vacuum to the neighborhood of another, then the
barrier crossing rate is a good measure of the topological transition rate.
Implicitly making this assumption, ref.~\cite{khlebnikov} claimed to show
that damping has no effect on topological transition rates,
in contradiction to the claims of ref.~\cite{arnold&mclerran}.
In this section, we
resolve this dilemma by showing that the assumed equality of the
barrier crossing and net topological transition rates fails if one studies
the full, short-distance theory.  The difference in the rates will turn
out to be precisely the suppression $\omega/g^2 T \sim g^2$ we have argued
arises from damping.


\subsection {Review of the microscopic origin of damping and the Langevin
             equation}

Since the purpose of this paper is pedagogy and not excruciating details
of formalism, we shall make some simplifications in order to elucidate
the physics involved.  First, we shall treat all the modes as classical
and simply assume there is an ultraviolet cut-off at momenta of order $T$
(which is where quantum mechanics enters to cut off the ultraviolet catastrophe
of classical thermal statistical mechanics).  Secondly, though we shall
consider all the hard modes of the theory, we shall only focus on the
one soft mode $\v a(\x)$ that is responsible for the decay of any particular
configuration on the barrier.  So we shall restrict
$\delta\A(t,\x) = \A(t,\x)-\Ab(\x)$
to
\begin {equation}
   \delta\A(t,x) \sim
      \delta A_{\rm soft} \; \v a(\x)
      + \int_{\rm hard} d^3q \> \delta\A(t,\q) \, e^{i\q\cdot\x}
   \,,
\label{eq:restrict}
\end {equation}
where the amplitude of $\v a(\x)$ is normalized to $O(1)$ near its
center.
Finally, imagine putting the system in a box so that we can discretize the
degrees of freedom.  Rather than studying gauge theory directly, we shall
begin by reviewing the derivation of damping for soft modes
in a generic theory with cubic interactions:
\begin {equation}
   \beta H = \half p_z^2 + \half K^2 z^2
     + \sum_i \left( |p_{y_i}|^2 + \Omega_i^2 |y_i|^2 \right)
     + z \sum_{ij} y_i^* \G_{ij} y_j
   \,.
\label{eq:model}
\end {equation}
Here $z$ represents the soft mode, $y_i$ the hard modes,
$K^2 \sim -(g^2 T)^2$ the curvature of the potential energy for
the unstable soft mode
[but we could also consider stable soft modes with $K^2 \sim +(g^2 T)^2$],
$\Omega_i$ the hard frequencies of
order $T$, and $\G$ the soft-hard-hard part of the
three-vector coupling.

The basic approximation is to realize that, provided the coupling $\G$ is
perturbative, the motion of the many hard modes is not much affected
by the motion of the soft mode.  To first approximation, then,%
\footnote{
  The simple time evolution (\protect\ref{eq:yapprox1})
  is to be understood as approximating
  the hard modes on time scales less than their thermalization time.
}
\begin {equation}
    y_i \simeq \alpha_i \, e^{-i\Omega_i t} \,,
\label{eq:yapprox1}
\end {equation}
where $\alpha_i$ are random phases and thermally-distributed random
amplitudes:
\begin {equation}
   \left\langle \alpha^*_i \alpha_j \right\rangle
   \equiv f(\Omega_i) \, \delta_{ij} = {1\over\Omega_i^2} \; \delta_{ij} \,,
\label{eq:alpha}
\end {equation}
which is the equipartition theorem with our normalizations
(\ref{eq:model}).
[With continuum normalizations, the quantity
corresponding to $f(\Omega_i)$ would be the classical limit
$T/\Omega_i$ of the Bose distribution $n({\Omega_i})$.]
Now consider the equation of motion for the mode of interest, $z$:
\begin {equation}
   \ddot z + K^2 z = - y^* {\cal G} y \,.
\label{eq:soft1}
\end {equation}
The leading approximation (\ref{eq:yapprox1}) produces an effectively random
force term $\xi(t)$ on the right-hand side,
\begin {equation}
   \xi(t) = - \sum_{ij} \alpha_i^* \G_{ij} \alpha_j \,
   e^{-i(\Omega_j-\Omega_i)t}
   \,.
\end {equation}
Using (\ref{eq:alpha}), the time correlation of this force is%
\footnote{
   Rigorously, the microcanonical problem involves {\it once} randomly
   choosing the $\alpha_i$ and then evolving the system, whereas
   (\ref{eq:alpha}) implies averaging over an ensemble of choices for
   each $\alpha_i$.  As long as the number of hard degrees of freedom that
   interact with the soft mode is large, there is no essential difference
   between these two cases.
}
\begin {equation}
   \left\langle \xi(t) \xi(t {+} \Delta t) \right\rangle =
      \sum_{ij} f(\Omega_i) f(\Omega_j) \, |\G_{ij}|^2 \,
      e^{-i(\Omega_j-\Omega_i)\Delta t}
   \,.
\label{eq:xixi}
\end {equation}
The force $\xi(t)$ is therefore a source of colored noise for the evolution
of the soft mode, and the power spectrum of that noise is given by the
Fourier transform of (\ref{eq:xixi}):
\begin {equation}
   \left\langle \xi(\omega) \xi(\omega') \right\rangle =
      2\pi\delta(\omega{-}\omega')
      \sum_{ij} f(\Omega_i) f(\Omega_j) \, |\G_{ij}|^2 \,
          2\pi\delta(\Omega_j{-}\Omega_i{-}\omega)
   \,.
\label{eq:xi power}
\end {equation}
The square matrix element just represents the process of fig.~\ref{fig:impi}
and, if one returns to continuum language, the power spectrum turns out to
be simply
\begin {equation}
   \left\langle \xi(\omega,\p) \xi(\omega',\p') \right\rangle =
      - 4\pi\delta(\omega{-}\omega') \, \delta^3(\p{-}\p') \,
      n(\omega) \, \Im\Pi(\omega)
   \,.
\label{eq:xi power2}
\end {equation}

So far we have reviewed the origin of an effectively random force term
in the equation of motion for soft modes (a term that we did not discuss
in our quick analysis of section 1).  To see the origin of damping, one
must consider the perturbative effect of the soft mode on the motion of
the hard modes.  The equations of motion for the $y_i$ are
\begin {equation}
   \ddot y_i + \Omega_i^2 \, y_i = - z(t) \sum_j \G_{ij} y_j \,.
\end {equation}
The perturbation to the solution is easily obtained by using the free
solution (\ref{eq:yapprox1}) for $y$ on the right-hand side
and then Fourier transforming:
\begin {equation}
   y_i(\omega) = \alpha_i \, 2\pi \delta(\omega{-}\Omega_i)
    + {1\over(\omega+i\epsilon)^2-\Omega_i^2}
      \sum_j {z(\omega {-} \Omega_j)} \G_{ij} \alpha_j
    + O(z^2)
   \,,
\label{eq:approx2}
\end {equation}
where we have used the retarded solution for the response of $y_i$ to $x$.
Putting this solution into the soft-mode equation (\ref{eq:soft1}) now
gives
\begin {equation}
   \bigl(-\omega^2 + K^2\bigr) z = \xi(\omega) +
       2 \sum_{ijk} \G_{ij} \G_{jk} \, \alpha_i^* \alpha_k \,
       {z(\omega {+} \Omega_i {-} \Omega_k)\over\omega^2-\Omega_j^2}
   + O(z^2)\,.
\end {equation}
Averaging the second term on the right-hand side over the random amplitudes
$\alpha_i$ finally yields
\begin {equation}
   \bigl(-\omega^2 + K^2 + \Pi(\omega)\bigr) z \simeq \xi(\omega)
   \,,
\end {equation}
where
\begin {equation}
   \Pi(\omega) =
       2 \sum_{ij} f(\Omega_i) {|\G_{ij}|^2 \over
          (\omega+\Omega_i+i\epsilon)^2-\Omega_j^2}
   \,.
\end {equation}
This is just the discretized version of the soft-mode self-energy of
fig.~\ref{fig:pi}.

The purpose of this review has been to show that the effective equation
of motion (\ref{eq:dA0}), which we first used to argue that the decay time is
$O(1/g^4 T)$, should more accurately be a Langevin equation of the
form
\begin {equation}
   (-\omega^2 + K^2 + \Pi(\omega)) \, \delta A_{\rm soft} = \xi(\omega)\,,
\end {equation}
where $\xi(\omega)$ is a random force with
$\langle\xi\rangle = 0$ and power spectrum
(\ref{eq:xi power2}).  For a stable soft mode ($K^2 > 0$), the decay of the
mode due to the imaginary part of the $\Pi(\omega)\delta A$ term and
the excitation of the mode by the random force term balance each other
to maintain thermal equilibrium of the soft mode.

Now (dropping the subscript ``soft'') consider the solution
\begin {equation}
   \delta A(t) = \delta \bar A(t) + \Delta A(t) \,,
\end {equation}
where $\delta\bar A(t)$ is the solution to the homogeneous equation that
we originally considered in section 1 and $\Delta A(t)$ are the fluctuations
induced by $\xi$:
\begin {equation}
   \Delta A(\omega) = {\xi(\omega) \over (-\omega^2 + K^2 + \Pi(\omega))} \,.
\end {equation}
Computing
the power spectrum of $\Delta A$
using (\ref{eq:xi power2}), one recovers
our earlier result of (\ref{eq:imD}) for the power $\rho_A$, projected onto the
soft mode under consideration.


\subsection {What does the decay look like?}

Let's now return to our characterization (\ref{eq:power}) of the
power of the fluctuations in $A$ integrated over frequencies of
order $\omega$.  We'd now like to investigate how much the relevant soft
mode oscillation $\Delta A(t)$ wiggles in time so that we can
assess whether the barrier is crossed once or multiple times per
net transition.

\begin {figure}
\vbox
   {%
   \bigskip
   \begin {center}
      \leavevmode
      
      \epsfbox [150 250 500 520] {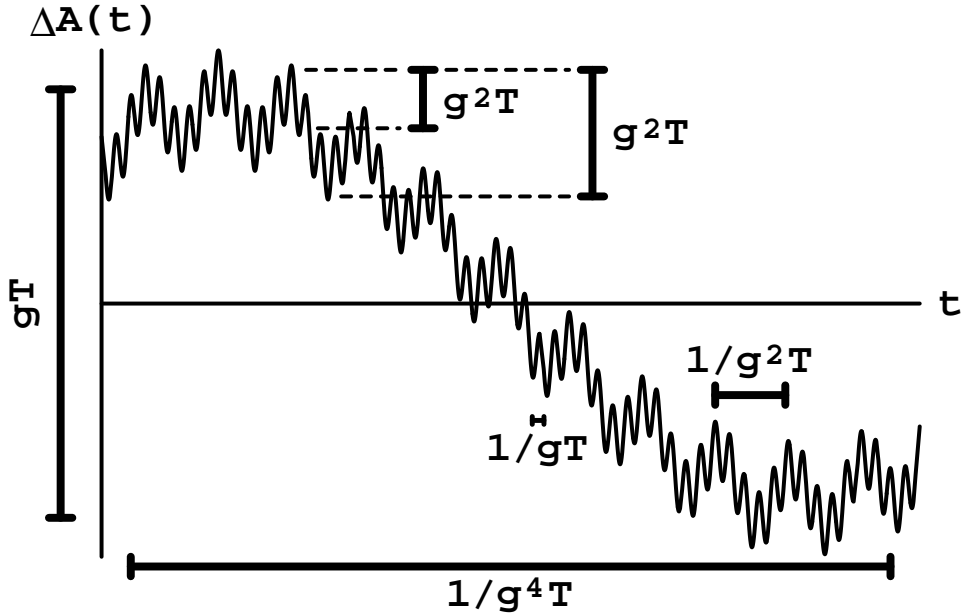}
   \end {center}
   \caption
       {%
       A schematic picture of the time evolution of the fluctuations
       $\Delta A(t)$ showing fluctuations of scales
       $\omega \sim g T$, $g^2 T$, and $g^4 T$.
       Keep in mind, however, that there is really a spectrum of
       fluctuations between $\omega \sim g^2 T$ and $\omega \sim g^4 T$
       whose amplitude grows as $\omega$ decreases.
       \label{fig:fluctuations}
       }%
   }%
\end {figure}

First, how many times does the motion of $\Delta A(t)$ change direction
per unit time?  This is equivalent to asking about the fluctuations
in $\Delta \dot A(t)$.  We can obtain the integrated power in
$\dot A$ simply by multiplying (\ref{eq:power}) by $\omega^2$:
\begin {equation}
   \omega \rho_{\dot A} \sim \cases{
       T                 \,,    &    $\omega \sim g T$ \,;              \cr
       \omega            \,,    &    $g^4 T \lesssim \omega \le p$ \,;  \cr
       \omega^3/g^8T^2   \,,    &    $\omega \lesssim g^4 T$ \,.        \cr
   }
\label{eq:powerdot}
\end {equation}
Unlike the power for the amplitude $A$ of fluctuations, the power for
$\dot A$ is dominated by plasma oscillations $\omega \sim g T$ and
not by low frequencies $\omega \sim g^4 T$.
The time scale for the motion of $\Delta A(t)$ to change direction is
therefore $1/gT$.  The power in $\Delta A$ and $\Delta \dot A$ can be
summarized by considering three characteristic frequency scales: $gT$, $g^2 T$,
and $g^4 T$.  The results of (\ref{eq:power}) and (\ref{eq:powerdot})
are summarized for these scales in
table~\ref{tab:fluctuations} and depicted
schematically in fig.~\ref{fig:fluctuations}.
Remember that the oscillations $\Delta A$
are superimposed on top of the slow net motion of the
homogeneous solution $\bar A$ for the damped decay.

\begin {table}
\begin {center}
\tabcolsep 10pt

\begin {tabular}{c|cc}
\hline
   $\omega$    &  $|A|$    &   $|\dot A|$   \\
\hline
   $gT$        &  $g^2 T$  &   $g^3 T^2$    \\
   $g^2 T$     &  $g^2 T$  &   $g^4 T^2$    \\ \ \ \ 
   $g^4 T$     &  $g   T$  &   $g^5 T^2$    \\
\hline
\end {tabular}
\end {center}
\caption
    {%
       Amplitude of fluctuations in $A(t)$ and $\dot A(t)$ corresponding
       to three characteristic frequency scales.
    }
\label {tab:fluctuations}
\end {table}

Now let us estimate how many times the system crosses $\delta A = 0$
during a topological transition.  The time scale for the net transition is
$t_{\rm net} \sim 1/g^4 T$, over which $\delta A$ changes magnitude by
$\delta_{\rm net} A \sim g T$.
During that time, $\delta A(t)$ changes directions of order
$g T t_{\rm net} \sim 1/g^3$ times
due to plasma oscillations whose amplitude is
$O(g^2 T)$.  The spectrum of fluctuations of the system may be rather
complicated, but the total time the system spends within $O(g^2 T)$ of
$\delta A = 0$ will be of order
\begin {equation}
   t_{\rm net} \times {g^2 T \over \delta_{\rm net} A}
   \sim
   {1 \over g^3 T} \,.
\end {equation}
During this time, the plasma fluctuations are large enough to drive
the system back and forth across $\delta A = 0$ with frequency
$g T$.  Thus, during one net transition, the system crosses
through $\delta A = 0$ of order $1/g^2$ times.  Any calculation which
simply computes the rate of barrier crossing per unit time will overestimate
the rate of net transitions by a factor of $1/g^2$.


\subsection {Being more precise about the ``barrier''}

In the previous section, we implied that crossing
$\delta A_{\rm soft} = 0$ many
times during a transition implies that the energy barrier hypersurface
is crossed many times during a transition.
However, $\delta A_{\rm soft}$
is just the projection of $\A-\Ab$ onto the soft mode
of interest, and the exact equation for the energy barrier hypersurface is not
really as simple as $\delta A_{\rm soft} = 0$.
Perhaps, as the hard modes oscillate,
the $\delta A_{\rm soft}$
location of the barrier also oscillates.  In order to confirm
our picture, we should check that the
effect of hard modes of the shape of the barrier is small enough not to
affect our argument.  We verify this in Appendix~\ref{apndx:barrier}.


\section {Implications for classical lattice thermal field theory}

The estimates of damping we have made have all been based on continuum,
quantum, thermal field theory, where the effective ultraviolet cut-off
for any thermal effects is $O(T)$, since the distribution of particles
with momenta $\gg T$ is Boltzmann suppressed.  One of the numerical testing
grounds used in the literature for studying topological transition rates,
however,
has been {\it classical} thermal field theory on a spatial lattice.  In
classical thermal field theory, modes of arbitrarily high momenta are
thermally excited (leading to the ultraviolet catastrophe of the classical
blackbody problem), and the ultraviolet cut-off is the inverse lattice
spacing $a^{-1}$ rather than $T$.  As we shall argue below, this infinite
growth in the number of relevant hard modes as $a\to0$ leads to
infinitely strong damping of the transition rate in the continuum limit
of classical thermal field theory.

   In the limit $\omega,p \ll T$, the usual continuum result for the
imaginary part of the transverse
self-energy $\PiT$ from figs.~\ref{fig:pi} and \ref{fig:impi}
is of the form
\begin {eqnarray}
   \Im\PiT(\omega,\p)
   &\sim& g^2 \int {d^3 q\over 2|\q|} \>
      n_{q} \left[|{\cal M_\omega}|^2 - |{\cal M}_{-\omega}|^2\right]
      \delta(\omega |\q| - \p\cdot\q)
\nonumber\\
   &\sim& g^2 \, {\omega\over p} \int_0^\infty dq \> q n_q
\nonumber\\
   &\sim& g^2 T^2 \, {\omega\over p} \,,
\label{eq:impi2}
\end {eqnarray}
where ${\cal M}_\omega$
is the vertex in the left-hand figure of fig.~\ref{fig:pi}.
The $q$ integral is dominated by the ultraviolet and cut-off by
$q \sim T$ for the quantum case, shown above.

For classical field theory, the only difference is that
\begin {equation}
   n_{\q} \to {T\over\omega_{\q}}
\end {equation}
and the ultraviolet cut-off is now the inverse lattice spacing $a^{-1}$,
so that the last step of (\ref{eq:impi2}) now gives
\begin {equation}
   \Im\Pi_{\rm cl}(\omega,p) \sim g^2 T a^{-1} \, {\omega\over p},
   \qquad (\omega \ll p) \,.
\end {equation}
Damping will therefore suppress the transition rate by an
extra factor of $a T$ compared to the quantum field theory
case, so that the rate is $O(\alpha^5 a T^5)$ and vanishes in the
$a\to0$ limit.

This result is in apparent contradiction with the numerical results of
ref.~\cite{ambjorn},
which claimed to find $O(\alpha^4 T^4)$ behavior for the rate.
Our result is suppressed by an additional factor of $O(\alpha a T)$,
which is one power of the dimensionless lattice coupling $\beta$.
Ref.~\cite{ambjorn} used values of $\beta$
ranging from 10 to 14, so they should find a 40\% violation of
the assumed $O(\alpha^4 T^4)$ scaling of the rate.
Such an effect is clearly inconsistent with their statistical
errors.  (See their fig.~3.)

There is a possible problem, however, with the assumption that
ref.~\cite{ambjorn}
is in fact measuring the topological transition rate.
The problem stems from
the subtleties of trying to define topological winding number on the lattice.
The idea in ref.~\cite{ambjorn} was
to measure a lattice analog of the square of
the topological winding:
\begin {equation}
   \left\langle
      \left(\int_0^t dt \int d^3 x \> \tr F \tilde F\right)^2
   \right\rangle
   = \left\langle \left(\Delta \Ncs(t) \right)^2 \right\rangle \,,
\label {eq:measure}
\end {equation}
where it is now convenient to consider field strengths normalized so that
the action is $F_{\mu\nu}^2/g^2$.
Under the picture of fig.~\ref{fig:ridge1},
the topological winding in pure gauge theory
should randomly diffuse away from zero, so that at large times
(\ref{eq:measure}) behaves like $\Gamma V t$ where $\Gamma$ is the transition
rate.  The authors implement this procedure on the lattice by making
a lattice approximation to $\tr F \tilde F$, which schematically has
the form of the cross-product
\begin {equation}
   \tr F \tilde F  \to  E^a \times \tr(U \sigma^a) \equiv J \,,
\label{eq:lattice}
\end {equation}
where $U$ represents plaquettes perpendicular to the links the 
electric field $E$
lives on.
(See ref.~\cite{ambjorn} for the detailed expression.)  The potential
problem with this representation is that, unlike the continuum
expression for $\tr F \tilde F$, the integral ({\it i.e.}, lattice sum) of $J$
over space is {\it not}
a total time derivative.  The time integral of this lattice analog
to $\int d^3 x \> \tr F \tilde F$ does not just depend on the initial
and final configurations but depends on the path used to get from one
to another.  In particular, consider a path in time that starts from
some configuration near the vacuum, never makes any non-perturbative
excursions from it (and so in particular never makes a topological transition),
and finally ends up back at the initial configuration.  The lattice
analog of the left-hand side of (\ref{eq:measure}),
$\left(\int dt \; d^3 x \> J\right)^2$,
would not be zero.
Perturbative fluctuations can therefore either increase or decrease
$\int dt \, d^3x \> J$ without increasing the energy,
and so there will be a purely perturbative contribution to the
diffusion.%
\footnote{
   This point has also been observed by Ambjorn and Krasnitz and will
   be addressed in a forthcoming publication.
}

To estimate the size of this lattice artifact in the diffusion
rate $\Gamma$, consider expanding (\ref{eq:lattice}) in powers of
the lattice spacing $a$ in lattice perturbation theory, remembering
that $U \sim e^{i a^2 B}$.  The leading $E\cdot B$ term is a total
time derivative and does not cause problems.  As an example
of a subleading term, consider a term involving $E$ and three powers
of $B$:
\begin {equation}
   \tr F \tilde F  \to  \tr F \tilde F  +  a^4 E B B B + \cdots
\end {equation}
Now consider the contribution of this term in the right-hand side of
(\ref{eq:measure}):
\begin {equation}
   a^8 \int_0^t dt_1 \int_0^t dt_2
   \left\langle \left(\int d^3 x \> E B B B\right)_{t_1}
                \left(\int d^3 x \> E B B B\right)_{t_2}
   \right\rangle
\end {equation}
The correlation is dominated by the ultraviolet and will be quasi-local in
space and time.  So, in the large time limit, it becomes
\begin {equation}
   a^8 V t \int dt' \, d^3 x
   \left\langle \left(E B B B\right)_{0,0}
                \left(E B B B\right)_{t',\x}
   \right\rangle
   \sim \beta^{-4} a^{-4} V t
   \sim g^8 T^4 V t
\end {equation}
This lattice artifact therefore gives a contribution to the
measured diffusion rate $\Gamma$ of $O(\alpha^4 T^4)$.  The moral
is that purely perturbative effects, having nothing to do with
true topological transitions, might obscure the true topological transition
rate, which we have argued is $O(\alpha^5 a T^5)$.


\bigskip

This work was supported by the U.S. Department of Energy,
grant DE-FG06-91ER40614.
We would like to thank Jan Ambj{\o}rn, Patrick Huet, Alex Krasnitz,
Larry McLerran, Misha Shaposhnikov,
and Frank Wilczek
for useful conversations (some of which date back
ten years).


\appendix

\section {Hard mode effects on the barrier surface shape}
\label{apndx:barrier}

  As discussed in section 1, the barrier hypersurface is the surface which
(1) separates the vacua and (2) maps into itself when the gradient of
the potential is followed.  In this appendix, we want to focus on
whether the hard modes have a significant impact on the shape of the
surface.  Specifically, consider (a) one soft, unstable mode of interest,
and (b) all hard modes,
as in the generic model of (\ref{eq:model})
with $K^2 = -\kappa^2 < 0$.
What is the equation of the barrier hypersurface in this subspace?

We shall restrict
attention to typical hard mode amplitudes in the thermal bath,
{\it i.e.}, $|y_i| \sim 1/\Omega_i$.  Secondly, we shall assume that
the interaction ${\cal G}$ between the soft mode and any {\it individual}
hard mode is perturbative, as it indeed is hot gauge theory.  The
translation of this condition to the generic model (\ref{eq:model}) is
easily made by considering stable soft modes, in which case $|z|$
is typically $1/\kappa$ and our perturbative condition is
$x y \G y \ll 1$, giving $\G \ll \kappa\Omega^2$.

Under these conditions, our result for the ridge equation is
\begin {equation}
   z = \sum_{ij} y_i^* {\G_{ij} \over \Omega_i^2 + \Omega_j^2} \, y_j
   \equiv \v y^* S \v y
   \,.
\label{eq:surface}
\end {equation}
This is easily checked as follows.  First, it contains
$(z,\v y, \v y^*) = (0,\v 0,\v 0)$
as it should.  Next, we must check that
\begin {equation}
   \nabla V(z,\v y^*,\v y)
   = (-\kappa^2 z + \v y^* \G \v y,
      \Omega^2 \v y + z \G \v y,
      \v y^* \Omega + z \v y^* G)
\end {equation}
lies in the tangent plane to the surface (within our approximations).
On the surface (\ref{eq:surface}),
\begin {equation}
   \nabla V(z,\v y^*,\v y) 
   \simeq (\v y^* \G \v y, \Omega^2 \v y, \v y^* \Omega^2) \,.
\label {eq:grad}
\end {equation}
The tangent plane to the surface is spanned by
\begin {eqnarray}
   {d\over d y_i^*} (z(\v y^*,\v y), \v y^*, \v y)
   = \left( (S \v y)_i, \ehat_i, 0
     \right) \,,
\\
   {d\over d y_i} (z(\v y^*,\v y), \v y^*, \v y)
   = \left( (S \v y)_i, 0, \ehat_i
     \right) \,,
\end {eqnarray}
and so
\begin {equation}
   n = ( 1 , - S \v y , - \v y^* S )
\label{eq:n}
\end {equation}
is normal to the surface.
(\ref{eq:grad}) and (\ref{eq:n}) then give $n^* \cdot \nabla V \simeq 0$
as desired.

To count the number of barrier crossings in section 3, we should really
have studied the evolution of $z - \v y^* S \v y$ rather than the
evolution of $z$.  We are now in a position to check whether
this makes any difference by checking the amplitude of the fluctuations
in $\v y^* S \v y$.  Using the leading-order behavior (\ref{eq:yapprox1})
for the $y_i$ and averaging over the amplitudes yields the equal time
amplitude
\begin {equation}
   \left\langle (\v y^* S \v y)^2 \right\rangle
   = \sum_{ij} f(\Omega_i) f(\Omega_j) |S_{ij}|^2 \,.
\label{eq:corr}
\end {equation}
Comparing this to
\begin {equation}
   \left\langle \xi^2 \right\rangle
   = \sum_{ij} f(\Omega_i) f(\Omega_j) |G_{ij}|^2 \,.
\end {equation}
shows that (\ref{eq:corr}) is of order $\langle\xi^2\rangle$ times the
typical size of $\Omega^{-4}$ (which is $T^{-4}$).
The continuum version is that the equation for the
barrier surface is
\begin {equation}
   \delta A = \chi(\A_{\rm hard}(t))
\end {equation}
where
\begin {equation}
   \langle\chi\rangle = 0,
\qquad
   \langle\chi^2\rangle
        \sim T^{-4} \langle\xi^2\rangle
\end {equation}
and $\xi$ here is understood to be projected onto the soft mode.  Using
(\ref{eq:xi power2}) and $p \sim g^2 T$,
\begin {equation}
   \langle\chi^2\rangle
        \sim T^{-4} p^3 \int d\omega \> n_\omega \, \Im\Pi(\omega)
        \sim T^{-4} p^3 \times g^2 T^3
        \sim (g^4 T)^2 \,.
\end {equation}
This $O(g^4T)$ magnitude
of the fluctuations in the location of the surface
is much smaller than the magnitude $g^2 T$ 
plasma oscillations described in section 3.
The shape of the surface therefore has no effect
on our previous discussion of crossing the barrier.



\begin {references}

\bibitem {kajantie}
   K. Kajantie, M. Laine, K. Rummukainen, and M. Shaposhnikov,
   CERN report CERN-TH/96-126, hep-ph/9605288.

\bibitem {elitzur}
   S. Elitzur,
   Phys.\ Rev.\ {\bf D12}, 3978 (1975).

\bibitem {arnold&mclerran}
   P. Arnold and L. McLerran,
   Phys.\ Rev.\ {\bf D36}, 581 (1987).

\bibitem{khlebnikov}
   S. Khlebnikov and M. Shaposhnikov,
   Nucl.\ Phys.\ {\bf B308}, 885 (1988).

\bibitem {ambjorn}
   J. Ambj{\o}rn and A. Krasnitz,
   Phys.\ Lett.\ {\bf B362}, 97 (1995).

\bibitem {christ}
   N. Christ,
   Phys.\ Rev.\ {\bf D21}, 1591 (1980).

\bibitem{3d reduction}
    T. Appelquist and R. Pisarski,
      {\sl Phys.\ Rev.} {\bf D23}, 2305 (1981);
    S. Nadkarni,
      {\sl Phys.\ Rev.} {\bf D27}, 917 (1983);
      {\sl Phys.\ Rev.} {\bf D38}, 3287 (1988);
      {\sl Phys.\ Rev.\ Lett.} {\bf 60}, 491 (1988);
    N. Landsman,
      {\sl Nucl.\ Phys.} {\bf B322}, 498 (1989);
    K. Farakos, K. Kajantie, M. Shaposhnikov,
      {\sl Nucl.\ Phys.} {\bf B425}, 67 (1994).

\bibitem{hard thermal loops}
    E. Braaten and R. Pisarski,
    Phys.\ Rev.\ {\bf D45}, 1827 (1992);
    Nucl.\ Phys. {\bf B337}, 569 (1990).

\bibitem {pi}
   H. Weldon,
     Phys.\ Rev.\ {\bf D26}, 1394 (1982);
   U. Heinz,
     Ann.\ Phys.\ (N.Y.) {\bf 161}, 48 (1985); {\bf 168}, 148 (1986).

\bibitem {fetter&walecka}
   A. Fetter and J. Walecka,
   {\sl Quantum theory of Many-Particle Systems}
   (McGraw-Hill: 1971).

\end {references}
\end {document}